\documentclass[aps,prc,floatfix,groupedaddress,showpacs,amsfonts,nofootinbib,twocolumn]{revtex4}
\usepackage{bm}
\usepackage{epsfig}
\usepackage{amsmath}
\usepackage{dcolumn}
\usepackage{slashed}
\usepackage{placeins}

\begin{document}

\title{Dirac oscillators and the relativistic $R$ matrix}

\author{J. Grineviciute}
\affiliation{Department of Physics, Western Michigan University, Kalamazoo, MI 49008}

\author{Dean Halderson}
\affiliation{Department of Physics, Western Michigan University, Kalamazoo, MI 49008}

\begin{abstract}
The Dirac oscillators are shown to be an excellent expansion basis for solutions of the Dirac equation by $R$-matrix techniques.  The combination of the Dirac oscillator and the $R$-matrix approach provides a convenient formalism for reactions as well as bound state problems.  The utility of the $R$-matrix approach is demonstrated in relativistic impulse approximation calculations where exchange terms can be calculated exactly, and scattering waves made orthogonal to bound state wave functions. \end{abstract}

\pacs{24.10.-i, 24.10.Jv, 25.40.Cm }

\maketitle

\section{Introduction}

In a previous paper \cite{Ha88} a calculable form of the $R$-matrix procedure was derived for the scattering of two particles obeying a relative Dirac equation.  The procedure was demonstrated by calculations of 33.5 MeV neutrons from a Woods-Saxon well.  An expansion basis consisted of the set of free-particle Dirac solutions whose upper components were zero at twice the $R$-matrix radius, $a_{c}$.  This basis did not provide the same stability seen in nonrelativistic calculations \cite{P75} and hence other basis function were sought.  Dirac oscillators give one that stability.  The combination of the $R$-matrix approach with Dirac oscillators provides a convenient formalism for coupled-channel reactions in which the binary breakup channels satisfy relative Dirac equations and also for bound state problems such as in relativistic mean field theory (RMFT).  

In this paper the procedure is applied to relativistic impulse approximation (RIA) calculations.  Here the technique allows one to calculate the exchange terms exactly and also to make the scattered wave orthogonal to the bound states obtained from RMFT.  Calculations are performed with the relativistic Love-Franey (RLF) model of Horowitz \cite{H85} for the two-nucleon $t$ matrix.  Cross sections, analyzing power, and spin rotation matrices are calculated for protons on $^{16}$O, $^{40}$Ca, and $^{90}$Zr.  The RLF model was constructed to have a small exchange amplitude, but even so, calculation of the exact exchange proved to be significant, leading to a conclusion that the RLF $t$ matrices with pseudoscalar coupling give better agreement with experimental results than previous determined.  The primary difference between exact calculation of the exchange and the conventional approximation appears in the matrix elements between negative energy states.  

\section{Dirac oscillator}

Dirac oscillators are obtained by adding a three-vector potential, linear in the radial coordinate, to the free-particle Dirac equation.  They were introduced by Ito et al. \cite{IM67}  and later revived by Moshinsky and Szczepaniak \cite{M89} where the name, Dirac oscillator, was applied due to the nonrelativistic reduction appearing as the ordinary harmonic oscillator with a strong spin-orbit interaction.  Their symmetry Lie algebra was discussed by Quesne and Moshinsky \cite{M90} and they were employed in calculations for wave packets in $3 + 1$ dimensions by Rozmej and Arvieu \cite{R99}.

The Hamiltonian for the Dirac oscillator is given by 
\begin{equation} 
\label{Eq1}
i\hbar \left( \partial \psi/\partial t \right) = \left[c \bm {\alpha} \cdot \left( \textbf{p} - i m \omega \textbf{r} \beta \right) + m c^{2} \beta\right] \psi = E \psi .
\end{equation}
Analytical solutions may be found by writing the wave function in two-component form
\begin{equation} 
\label{Eq2}
\psi =
\begin{pmatrix}
{\left[F\left(r\right)/r\right]\Phi_{\kappa m}}\\
{\left[iG\left(r\right)/r\right]\Phi_{- \kappa m}}
\end{pmatrix}
,
\end{equation}
where
\begin{equation}
\label{Eq3}
\Phi_{\kappa m} = \sum_{m_{l} m_{s}} {C^{l\ 1/2\ j}_{m_{l} m_{s} m}
Y_{l m_{l}}\left(\theta , \phi \right)\chi_{m_{s}} }   \; ,
\end{equation}
$j=\left|\kappa\right|-1/2$, and $l=\kappa$ for $\kappa>0$, but $l=-\left(\kappa+1\right)$ for $\kappa<0$.  Insertion of Eq. (\ref{Eq2}) into Eq. (\ref{Eq1}) and projecting on the appropriate $\Phi_{\kappa m}$ yields two coupled equations for $F$ and $G$,
\begin{align}
\label{Eq4}
\left(E - m\right){F(r)}= & \left[-dG(r)/dr\right. \nonumber\\
& \left.+ \eta \left(j+1/2\right)  G(r) + m \omega r G(r)\right] 
\end{align}
and
\begin{align}
\label{Eq5}
 \left(E + m\right){G(r)} = & \left[ dF(r)/dr  \right. \; , \nonumber\\
& \left. + \eta \left(j+1/2\right) {F(r)} + m \omega r F(r) \right] \; ,
\end{align}
where $\eta=+1$ for $\kappa>0$ and $\eta=-1$ for $\kappa<0$. With $\hbar = c=1$ and $\alpha=\left(m\omega\right)^{1/2}$, Eqs. (\ref{Eq4}) and (\ref{Eq5}) have solutions
\begin{align}
\label{Eq6}
F\left(r\right) = e^{-{x/2}} x^{\left(l + 1\right)/2} L^{l + 1/2}_{n} \left(x\right) \ {\rm and} \nonumber\\
G\left(r\right) = - \frac{2 \alpha}{E + m} e^{-{x/2}} x^{\left(l^{\prime} + 1\right)/2} L^{l^{\prime} + 3/2}_{n^{\prime}} \left(x\right) 
\end{align}
for $\kappa<0$ and
\begin{align}
\label{Eq7}
F\left(r\right) = \frac{2 \alpha}{E - m} e^{-{x/2}} x^{\left(l + 1\right)/2} L^{l + 1/2}_{n} \left(x\right)  \ {\rm and}  \nonumber\\
G\left(r\right) = e^{-{x/2}} x^{\left(l^{\prime} + 1\right)/2} L^{l^{\prime} + 1/2}_{n} \left(x\right)
\end{align}
for $\kappa>0$, where $x=m\omega r^{2}$ and $L^{a}_{b}\left(x\right)$ is an associated Laguerre polynomial. In Eq. (\ref{Eq6}) $l^{\prime}=l+1$ and $n^{\prime}=n-1$. In (\ref{Eq7}) $l^{\prime}=l-1$ and $n^{\prime}=n$, with $n$ starting at zero. The energies are given by  $E^{2} = m^{2} + 4\left(n + l + 1/2\right)mw$ for $\kappa > 0$ and $E^{2} = m^{2} + 4nmw$ for $\kappa < 0$. 

These solutions can be normalized and placed in a form involving the ordinary, normalized radial harmonic oscillator wave functions, $u_{nl}\left(r\right)$. Eq. (\ref{Eq6}) becomes
\begin{align}
\label{Eq8}
&F\left(r\right) = u_{nl}\left(r\right)/\sqrt{1+A^{2}} \ {\rm and} \nonumber\\
& G\left(r\right) = u_{n^{\prime}l^{\prime}}\left(r\right) A/ \sqrt{1+A^{2}} ,
\end{align}
where
\begin{equation}
\label{Eq9}
A=-\frac{2 \alpha}{E + m} \sqrt{n} ,
\end{equation}
and Eq. (\ref{Eq7}) becomes
\begin{align}
\label{Eq10}
& F\left(r\right) = u_{nl}\left(r\right)/\sqrt{1+A^{2}} \ {\rm and} \nonumber\\
& G\left(r\right) = u_{n^{\prime}l^{\prime}}\left(r\right) A/ \sqrt{1+A^{2}} ,
\end{align}
where
\begin{equation}
\label{Eq11}
A=\left(E - m\right) \frac{1}{2 \alpha} \sqrt{\frac{n + l + 1}{2 \left(n + l\right) + 1}} ,
\end{equation}
A curiosity of the Dirac oscillators is that $G(r) = 0$ for $\kappa < 0$ and $n = 0$ and the negative energy state for $\kappa < 0$ corresponding to $n = 0$ does not exist. One needs both positive and negative energy solutions in the $R$-matrix expansion.

\section{The R matrix}

The R-matrix for the Dirac equation was developed in Ref. \cite{Ha88} by following the same steps that were followed in the nonrelativistic case.  For the one-channel case the wave function is in the form of Eq. (\ref{Eq2}) and is expanded within the channel radius as $\Psi = \sum_{\lambda} A_{\lambda} \left|\lambda\right\rangle$. The set of $\left|\lambda\right\rangle$ will be Dirac oscillators.  The appropriate Block operator, whose purpose is described in Lane and Robson \cite{LaRo}, is constructed to be
\begin{equation}
\label{Eq12}
L\left(b\right) = \delta\left(r - a_{c}\right)
\begin{pmatrix}
{-b}&{i\left( {\bm {\sigma}}\cdot{\textbf{r}} \right)/r}\\
{0}&{0}
\end{pmatrix}
\end{equation}
and added to the Hamiltonian.  The natural choice of the boundary condition parameter, $b=G\left(a_{c}\right) / F\left(a_{c}\right)$, requires that $L\left(b\right) \Psi = 0$ and, therefore, $\left[H+L\left(b\right)\right]\Psi = E\Psi$. Insertion of the expansion into this equation gives
\begin{equation} 
\label{Eq13}
\sum_{\lambda^{\prime}} {\left[\left\langle \lambda\left| H - E \right| \lambda^{\prime}\right\rangle +  {\gamma_{\lambda} \left(b_{\lambda^{\prime}} - b\right) \gamma_{\lambda^{\prime}}} \right] A_{\lambda^{\prime}}} = 0 \: .
\end{equation}
where $b_{\lambda}=G_{\lambda}\left(a_{c}\right) / F_{\lambda}\left(a_{c}\right)$ and $\gamma_{\lambda}=F_{\lambda}\left(a_{c}\right)$. One can see that the lower component in the Dirac theory is taking the place of the derivative of the wave function in the nonrelativistic theory. 

The theory is placed in calculable form in the method of Philpott \cite{P75} in which one finds a transformation, $T$, such that 
\begin{equation}
\label{Eq14}
\sum_{\lambda\lambda^{\prime}} {T_{\lambda\mu} \left[ \left\langle \lambda\left| H \right| \lambda^{\prime}\right\rangle +   {\gamma_{\lambda} b_{\lambda^{\prime}} \gamma_{\lambda^{\prime}}} \right] T_{\lambda^{\prime} \mu^{\prime}}} = E_{\mu} \delta_{\mu\mu^{\prime}} .
\end{equation}
With this transformation, Eq. (\ref{Eq13}) becomes
\begin{equation}
\label{Eq15}
\sum_{\mu^{\prime}} {\left[ \left( E_{\mu} - E\right )\delta_{\mu\mu^{\prime}} -  {\gamma_{\mu} b \gamma_{\mu^{\prime}}} \right] A_{\mu^{\prime}}} = 0 ,
\end{equation}
where $\gamma_{\mu} = \sum_{\lambda} \gamma_{\lambda} T_{\lambda\mu}$ and $ A_{\mu} = \sum_{\lambda} T_{\lambda\mu}A_{\lambda}$. Equation (\ref{Eq15}) is solved for the $A_{\mu}$, $A_{\mu} = \gamma_{\mu} G\left(a_{c}\right) /\left(E_{\mu}-E\right) $, and these reinserted into Eq. (\ref{Eq15}) to give the $R$-matrix equation $\left(1 - bR\right)\gamma = 0$ or $R = F\left(a_{c}\right) / G\left(a_{c}\right)$, where
\begin{equation}
\label{Eq16}
R = \sum_{\mu} \gamma^{2}_{\mu} / \left(E_{\mu} - E\right) \; .
\end{equation}

\section{SCATTERING AMPLITUDES}

The regular and irregular Dirac Coulomb functions are generated as given by Young and Norrington \cite{YN94} employing the code COULCC\cite{T85}.  The upper (lower) component of the regular function will be specified by $F_{F}(G_{F})$, while the upper (lower) component of the irregular function will be specified by $F_{G}(G_{G})$.  The function $F_{F}\left(r\right) \rightarrow \sin \varphi \left(r\right)$ and $F_{G}\left(r\right) \rightarrow \cos \varphi \left(r\right)$, where $\varphi \left(r\right) = kr + y \ln 2kr - l\pi/2 + \delta^{\prime}_{\kappa}$, $k$ is the momentum of the proton in the center-of-momentum system, $y=Ze^{2}E/k$, $E^{2} = m^{2}_{p} + k^{2}$,
\begin{equation} 
\delta^{\prime}_{\kappa} = \Psi - \arg\Gamma\left({\gamma + iy}\right) + {\frac{\pi}{2}} {\left(l+1-\gamma\right)} \; , \nonumber
\end{equation}
\begin{equation} 
e^{2i\Psi} = \frac{ie^{2}Z/k-\kappa}{\gamma+iy} \; , \nonumber
\end{equation}
and $\gamma = \left(\kappa^{2}-Z^{2} e^{4}\right)^{\frac{1}{2}}$.  One constructs the incoming, $F_{I} = F_{G} - iF_{F}$, and outgoing, $F_{O} = F_{G} + iF_{F}$, solutions and defines the collision matrix from $F\rightarrow F_{I}-SF_{O}$. The relation $R = F\left(a_{c}\right) / G\left(a_{c}\right)$ gives
\begin{equation}
\label{Eq17} 
R=\left(F_{I}-SF_{O}\right)/\left(G_{I}-SG_{O}\right)  ,
\end{equation}
and therefore the collision matrix is determined for each $\kappa$ by
\begin{equation}
\label{Eq18} 
S=\left[F_{I}\left(a_{c}\right)/F_{O}\left(a_{c}\right)\right] \left[\left(1-L_{I}R\right)/\left(1-L_{O}R\right)\right] ,
\end{equation}
where $L_{I}=G_{I}\left(a_{c}\right)/F_{I}\left(a_{c}\right)$ and $L_{O}=G_{O}\left(a_{c}\right)/F_{O}\left(a_{c}\right)$.

\begin{figure}[!htbp]
\includegraphics[width=8cm]{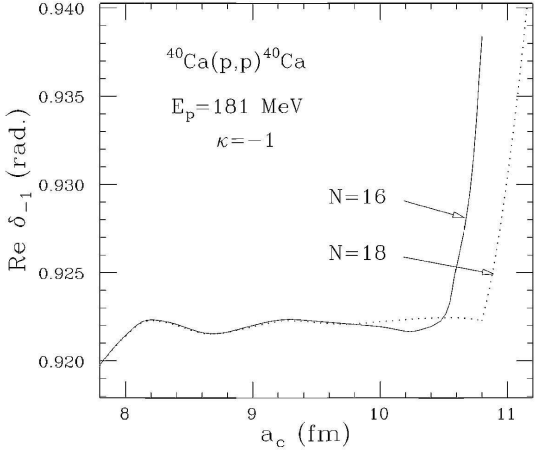}
\caption{\label{Fig1} The real part of the $\kappa = -1$ phase shift  as a function of $R$-matrix radius for $N=16$ and $18$. The calculation is for the phenomenological optical potential of Ref. \cite{A81}.}
\end{figure}

In practice one must chose an $R$-matrix radius and number of basis states for each $\kappa$.  The choice is made by looking at the phase shift as a function of $a_{c}$.  One finds a region where the phase shift is not changing.  This is illustrated in Fig. \ref{Fig1} for protons on $^{40}$Ca at 181 MeV with the phenomenological potential of Ref. \cite{A81}.  The curves are for a maximum value $(N)$ of $n$ set to 16 and 18.  One sees a range of about 2 fm where the real part of the phase shift is accurate to three figures.  In general the curves have a wider flat range and the flat range moves to larger radii as the number of oscillators increases.  Also, fewer oscillators are required for larger values of $l$.  Fixing the radii for each $\kappa$ can usually be accomplished by finding a radius for a sufficient number of oscillators $(N_{0})$ for $\kappa=-1$ and then keeping that radius for all $\kappa$ but changing the number of oscillators according to $N=N_{0}-0.3l$.

The Coulomb scattering amplitudes are given by 
\begin {align}
\label{Eq19} 
F_{C}  = \frac{1}{2ip} \sum^{\infty}_{l=0} & \left[ \left(l + 1\right)\left(e^{2i\delta^{'}_{-l - 1}} - 1\right) + l\left(e^{2i\delta^{'}_{l}} - 1\right) \right] \nonumber\\
& \times P_{l}\left(\cos\theta\right)  ,
\end{align}
\begin {equation}
G_{C}  = \frac{1}{2p} \sum^{\infty}_{l=0} \left[ e^{2i\delta^{'}_{-l - 1}} - e^{2i\delta^{'}_{l}} \right] P^{1}_{l}\left(\cos\theta\right) . \nonumber
\end{equation}
The series is summed using the reduction method where the amplitudes are multiplied by a function which vanishes at $\theta=0$ and then expanded in a series of Legendre polynomials:
\begin{align}
& \left(1  - \cos\theta\right)^{m} F_{c}   \nonumber\\
& = \sum_{l} a^{(m)}_{l} P_{l} (\cos\theta)  = \sum_{l} \left( a^{(m - 1)}_{l} - \frac{l}{2l - 1} a^{(m - 1)}_{l - 1} \right.  \nonumber\\
& \left. - \frac{l + 1}{2l + 3} a^{(m - 1)}_{l + 1} \right) P_{l} (\cos\theta) , \nonumber
\end{align}
\begin{align}
& \left(1 - \cos\theta\right)^{m} G_{c} \nonumber\\
& = \sum_{l} a^{(m)}_{l} P^{1}_{l} (\cos\theta)  = \sum_{l} \left(a^{(m - 1)}_{l} - \frac{l-1}{2l - 1} a^{(m - 1)}_{l - 1} \right.  \nonumber\\
& \left. - \frac{l + 2}{2l + 3} a^{(m - 1)}_{l + 1} \right) P^{1}_{l} (\cos\theta) . \nonumber
\end{align}
The nuclear scattering amplitudes are the obtained by replacing the Coulomb $t$ matrix, $t_{C}=i\left(1-e^{2i\delta^{\prime}_{\kappa}}\right)/2$, by $e^{2i\delta^{\prime}_{\kappa}} t_{N}$ in Eq. (\ref{Eq19}), where $t_{N}=i\left(1-S\right)/2$. The total scattering amplitude then becomes the sum nuclear plus Coulomb.

\section{IMPULSE APPROXIMATION}

As an example of combining the Dirac oscillators with the $R$-matrix techniques, calculations have been made with the RFL amplitudes of Horowitz.  These amplitudes have been employed in the previous RIA calculation of Murdock and Horowitz \cite{M87}.  The present calculation follows the procedures of Ref. \cite{M87}, but deviates by calculating the exchange term without the nuclear matter approximation for the exchange density.  As given in Ref. \cite{M87,HMS91}, one can write the RIA optical potential, acting on state $U_{0}$, for $L-S$ closed shells as
\begin {align}
\label{Eq20} 
& \left\langle \textbf{x}\left|V_{opt}\right| U_{0}\right\rangle \nonumber\\
& = - \frac{4\pi ik}{m_{p}} \sum_{L} \int {d^{3}x^{\prime}} \rho^{L} \left(\textbf{x}^{\prime}, \textbf{x}^{\prime} \right) t^{L}_{D} \left(\left|\textbf{x}^{\prime} -\textbf{x} \right| ; E\right) \lambda^{L} U_{0} \left(\textbf{x}\right) \nonumber\\
&- \frac{4\pi ik}{m_{p}} \sum_{L} \int {d^{3}x^{\prime}} \rho^{L} \left(\textbf{x}^{\prime}, \textbf{x} \right) t^{L}_{X} \left(\left|\textbf{x}^{\prime} -\textbf{x} \right| ; E\right) \lambda^{L} U_{0} \left(\textbf{x}^{\prime}\right) \; ,
\end{align}
where
\begin {equation} 
\label{Eq21} 
t^{L}_{Y} \left(\left|\textbf{x}\right|;E\right) \equiv \int \frac{d^{3}q}{\left(2\pi\right)^{3}} t^{L}_{Y} \left(\textbf{q},E\right) e^{i\textbf{q}\cdot \textbf{x}} ,
\end{equation}
\begin {equation}
\label{Eq22} 
\rho^{L} \left(\textbf{x}^{\prime},\textbf{x}\right) \equiv \sum^{occ}_{\alpha} \bar{U}_{\alpha} \left(\textbf{x}^{\prime}\right) \lambda^{L} U_{\alpha} \left(\textbf{x}\right) ,
\end{equation}
$\lambda^{L}=1$ or $\gamma^{0}$ for the scalar or vector potential, and $Y=D$ for direct or $Y=X$ for exchange. The exchange potential is, therefore, non-local, but can be made local by approximating $\rho^{L} \left(\textbf{x}^{\prime},\textbf{x}\right)$ in the method of Brieva and Rook \cite{BR77} as was done in Refs. \cite{M87,HMS91}.

However, it is possible to calculate the exchange exactly in the R-matrix approach.  Here one is calculating the matrix elements appearing in Eq. (\ref{Eq13}).  They take the form
\begin {align}
\label{Eq23} 
& \left\langle \gamma_{0} \psi_{nlj} \left| V_{exch} \right| \psi_{n^{\prime}l^{\prime}j^{\prime}} \right\rangle \nonumber\\
& = - \frac{4\pi ik}{m_{p}} \int d^{3}x^{\prime} d^{3}x \bar{\psi}_{nlj}\left(\textbf{x}\right) \rho^{L} \left(\textbf{x}^{\prime},\textbf{x}\right) \nonumber\\
& \times t^{L}_{X} \left(\left|\textbf{x}^{\prime} -\textbf{x} \right| ; E\right) \lambda^{L} \psi_{n^{\prime}l^{\prime}j^{\prime}} \left(\textbf{x}^{\prime}\right) ,
\end{align}
where $t^{L}_{X} \left(\left|\textbf{x}^{\prime} -\textbf{x} \right| ; E\right)$ is replaced by the integral form in Eq. (\ref{Eq21}).  The exponential is then expanded as 
\begin {align}
\label{Eq24} 
& e^{-i \textbf{q}\cdot \left(\textbf{x} - \textbf{x}^{\prime}\right)} = \left(4\pi\right)^{2} \sum_{LML^{\prime}M^{\prime}} \left[i^{L}j_{L}\left(qx^{\prime}\right)Y^{*}_{LM}\left(\hat{q}\right)Y_{LM}\left(\hat{x}^{\prime}\right)\right]  \nonumber\\
& \left(-i\right)^{L^{\prime}}j_{L^{\prime}}\left(qx\right)Y^{*}_{L^{\prime}M^{\prime}}\left(\hat{x}\right)Y_{L^{\prime}M^{\prime}}\left(\hat{q}\right)  .
\end{align}
The matrix element becomes 
\begin{align}
\label{Eq25} 
& \left\langle \gamma_{0} \psi_{nlj} \left| V_{exch} \right| \psi_{n^{\prime}l^{\prime}j^{\prime}} \right\rangle  \nonumber\\
&= - \frac{8ik}{m_{p}\left(2j + 1\right)} \sum_{j_{\alpha} \tau_{\alpha} L} \int q^{2}dq t^{\tau_{\alpha}}_{X} \left(q\right)\left( R^{FL}_{n\alpha} \left\langle j_{\alpha}l_{\alpha} \left\|Y_{L}\right\| jl\right\rangle \right. \nonumber\\
& \left. \pm R^{GL}_{n\alpha} \left\langle j_{\alpha} \bar{l}_{\alpha} \left\|Y_{L}\right\| j\bar{l}\right\rangle \right) \left(  R^{FL}_{\alpha n^{\prime}} \left\langle j_{\alpha}l_{\alpha} \left\|Y_{L}\right\| jl\right\rangle \right. \nonumber\\
& \left. \pm R^{GL}_{\alpha n^{\prime}} \left\langle j_{\alpha} \bar{l}_{\alpha} \left\|Y_{L}\right\| j\bar{l}\right\rangle \right) \; ,  
\end{align}
where the plus (minus) signs are for the vector (scalar) potential, $R^{FL}_{n \alpha} = \int dx F_{nlj} \left(x\right) j_{L}\left(x\right) F_{\alpha}\left(x\right)$,  $R^{GL}_{n \alpha} = \int dx G_{n\bar{l}j} \left(x\right) j_{L}\left(x\right) G_{\alpha}\left(x\right)$, $\bar{l} = l \pm 1$ for$ j = l\mp 1/2 $, and the reduced matrix elements are as defined in de-Shalit amd Talmi \cite{ShTa}.

\section{Results}

\begin{figure*}[!htbp]
\includegraphics[width=17cm]{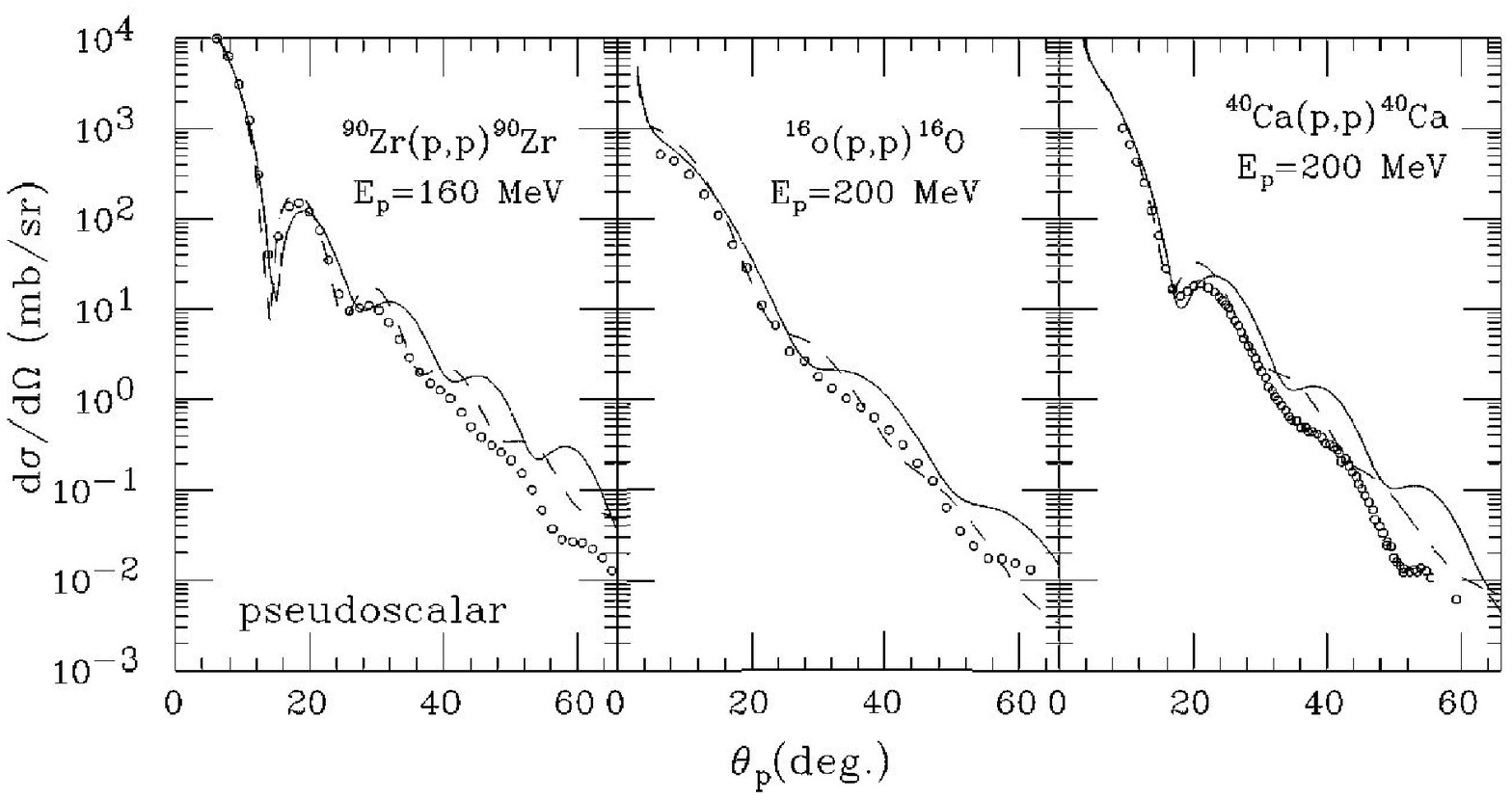}
\caption{\label{Fig2} Elastic scattering cross sections. Solid (dashed) lines are from calculations with approximate (exact) exchange and pseudoscalar  coupling. Data are from Ref. \cite{S85} and \cite{S82} as given in \cite{HMS91}.}
\end{figure*} 
\begin{figure*}[!htbp]
\includegraphics[width=17cm]{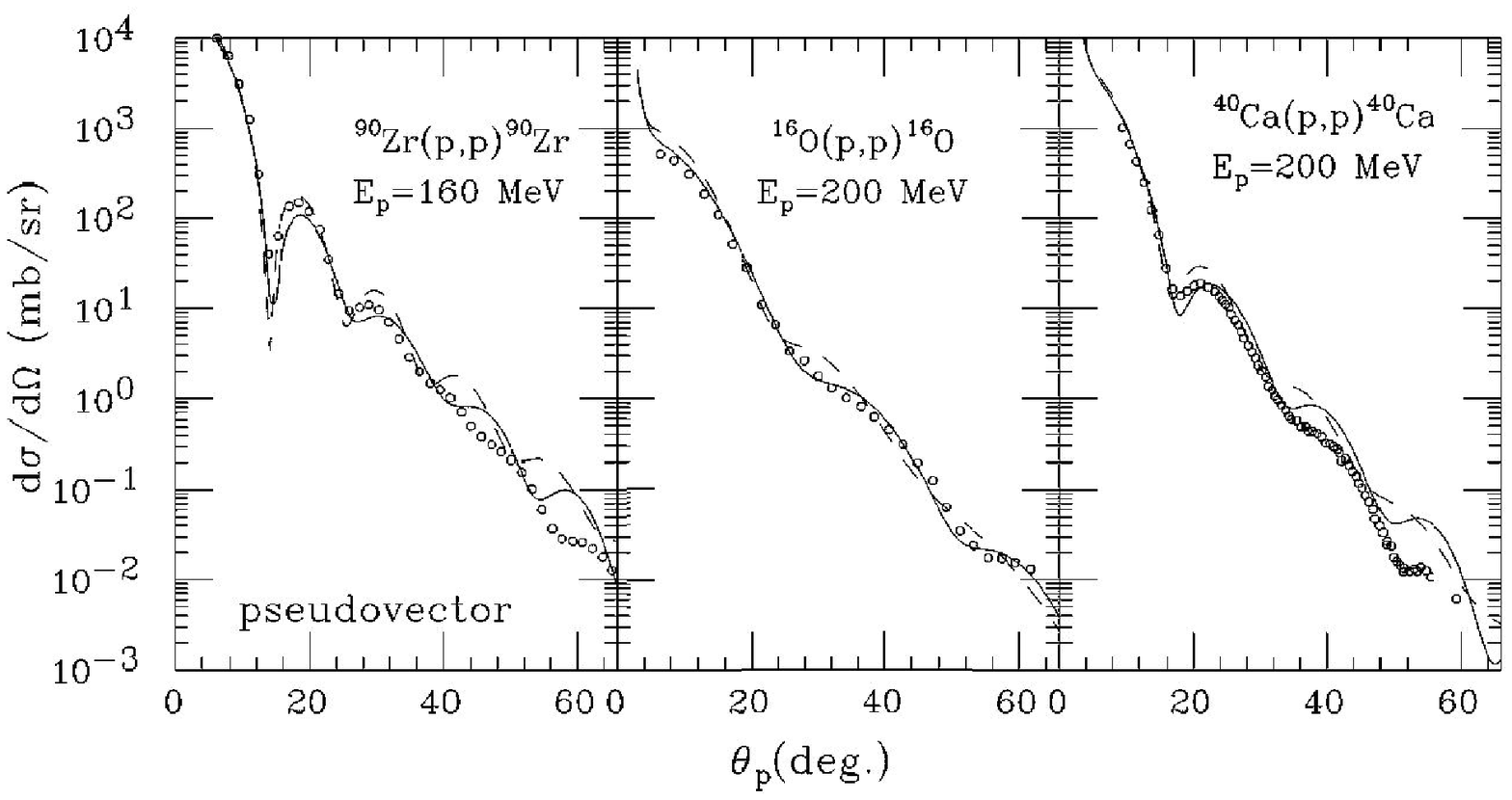}
\caption{\label{Fig3} Elastic scattering cross sections. Solid (dashed) lines are from calculations with approximate (exact) exchange and pseudovector  coupling. Data are from Ref. \cite{S85} and \cite{S82} as given in \cite{M87}.}
\end{figure*}
\begin{figure*}[!htbp]
\includegraphics[width=17cm]{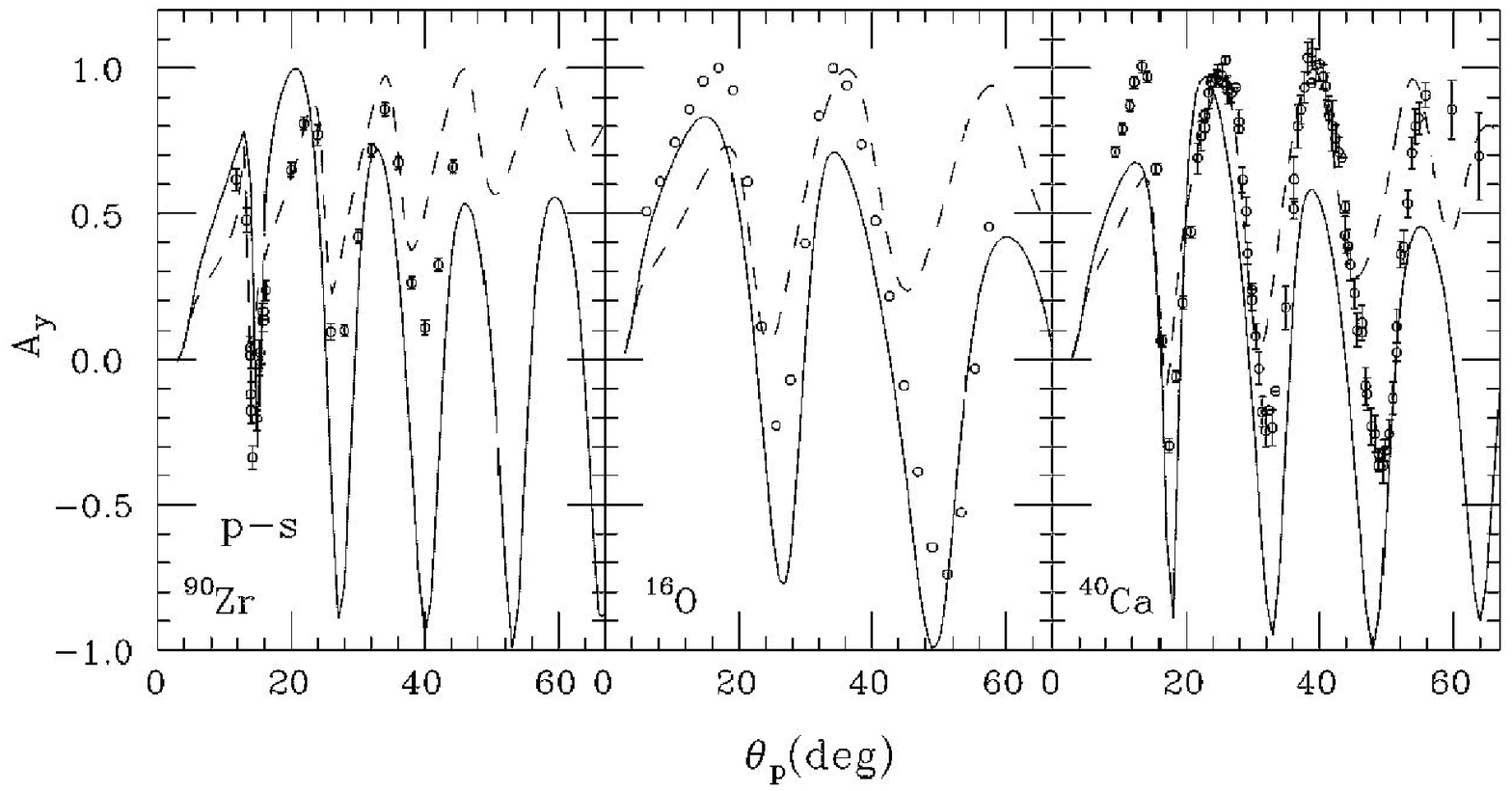}
\caption{\label{Fig4} Analyzing power. Solid (dashed) lines are from calculations with approximate (exact) exchange and pseudoscalar  coupling. Data are from Ref. \cite{S85} and \cite{S82} as given in \cite{M87}.}
\end{figure*}
\begin{figure*}[!htbp]
\includegraphics[width=17cm]{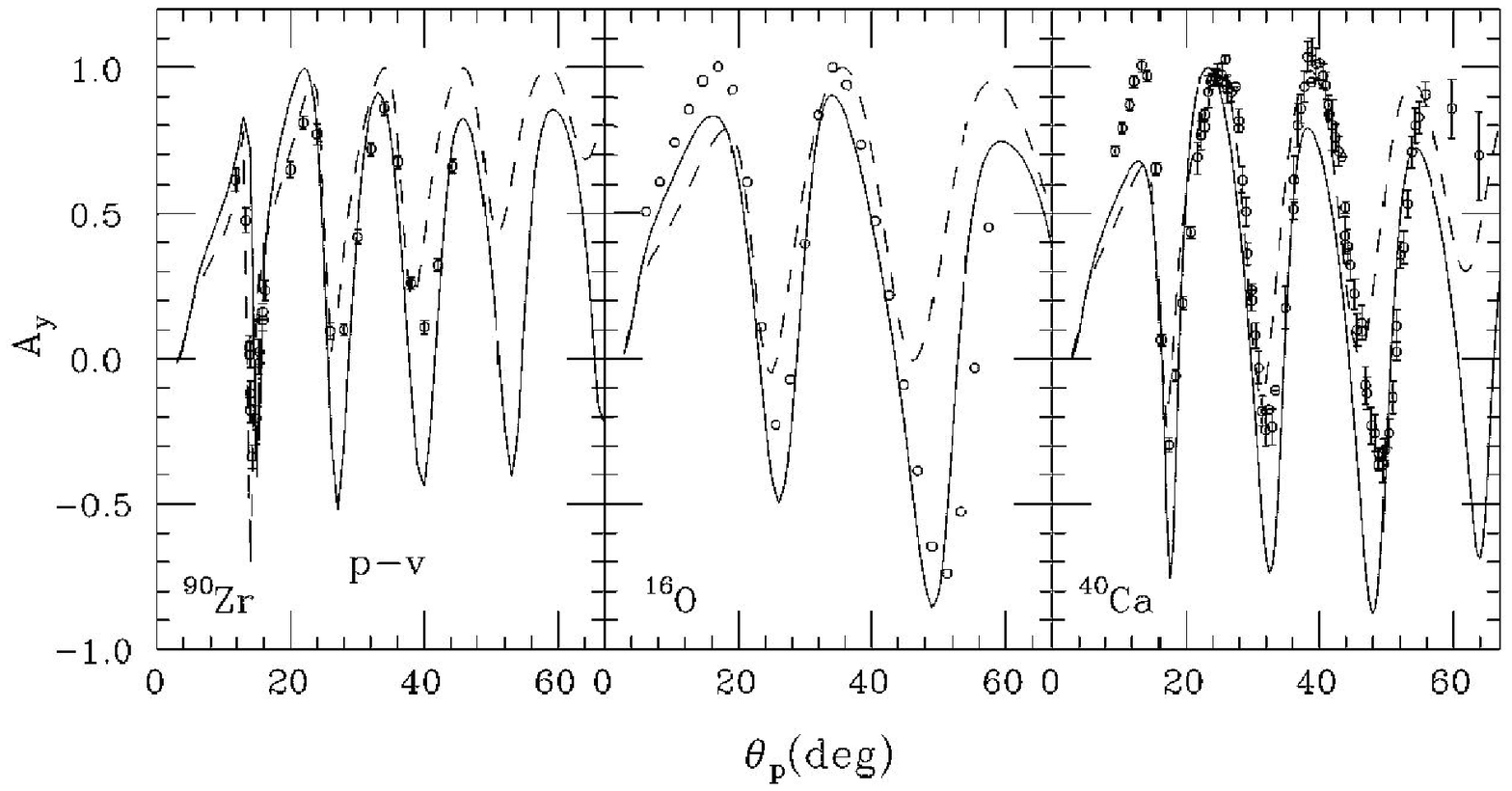}
\caption{\label{Fig5} Analyzing power. Solid (dashed) lines are from calculations with approximate (exact) exchange and pseudovector  coupling. Data are from Ref. \cite{S85} and \cite{S82} as given in \cite{M87}.}
\end{figure*}
\begin{figure*}[!htbp]
\includegraphics[width=17cm]{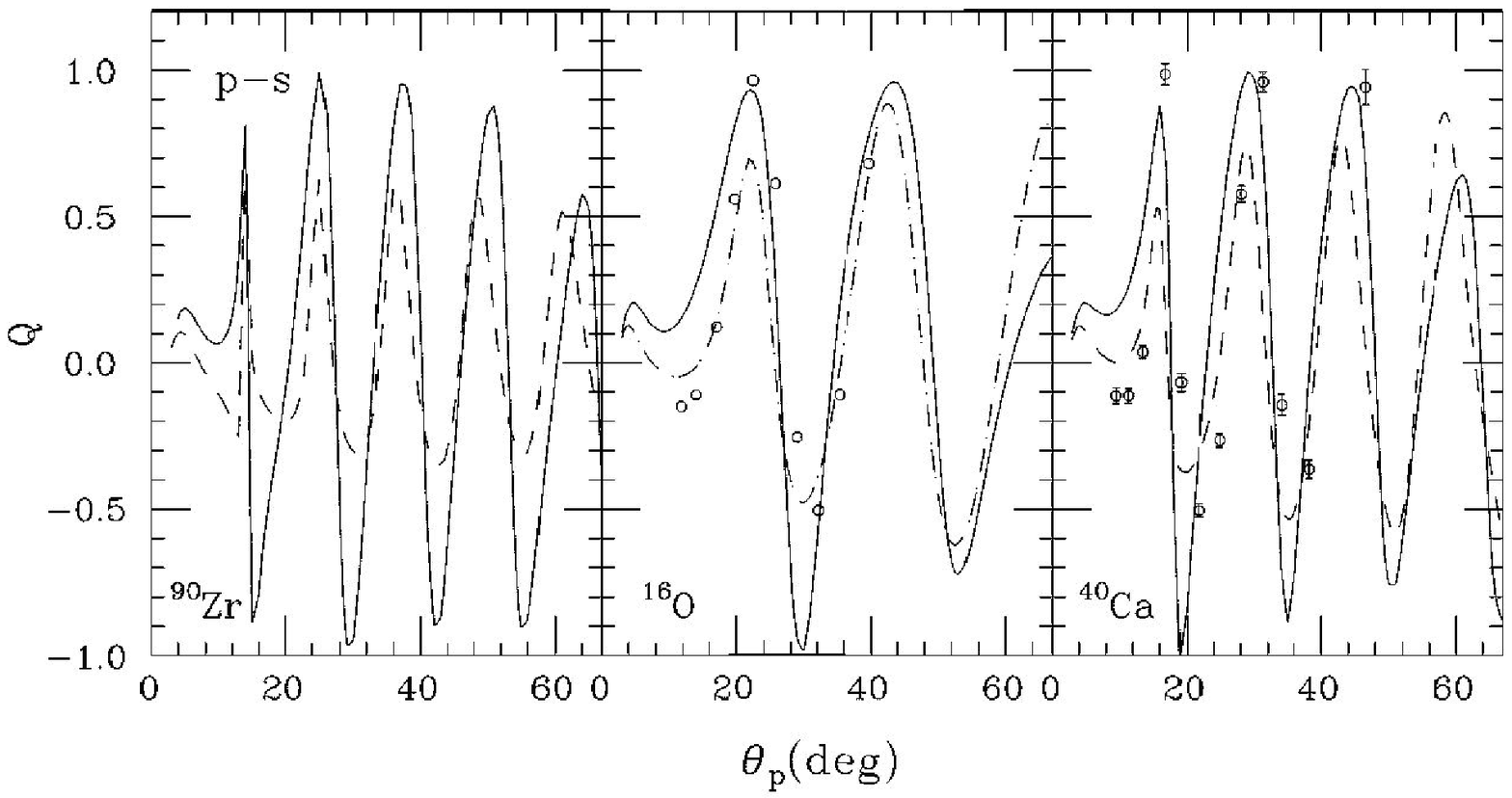}
\caption{\label{Fig6} Spin rotation matrix. Solid (dashed) lines are from calculations with approximate (exact) exchange and pseudoscalar  coupling. Data are from Ref. \cite{S85} and \cite{S82} as given in \cite{M87}.}
\end{figure*}
\begin{figure*}[!htbp]
\includegraphics[width=17cm]{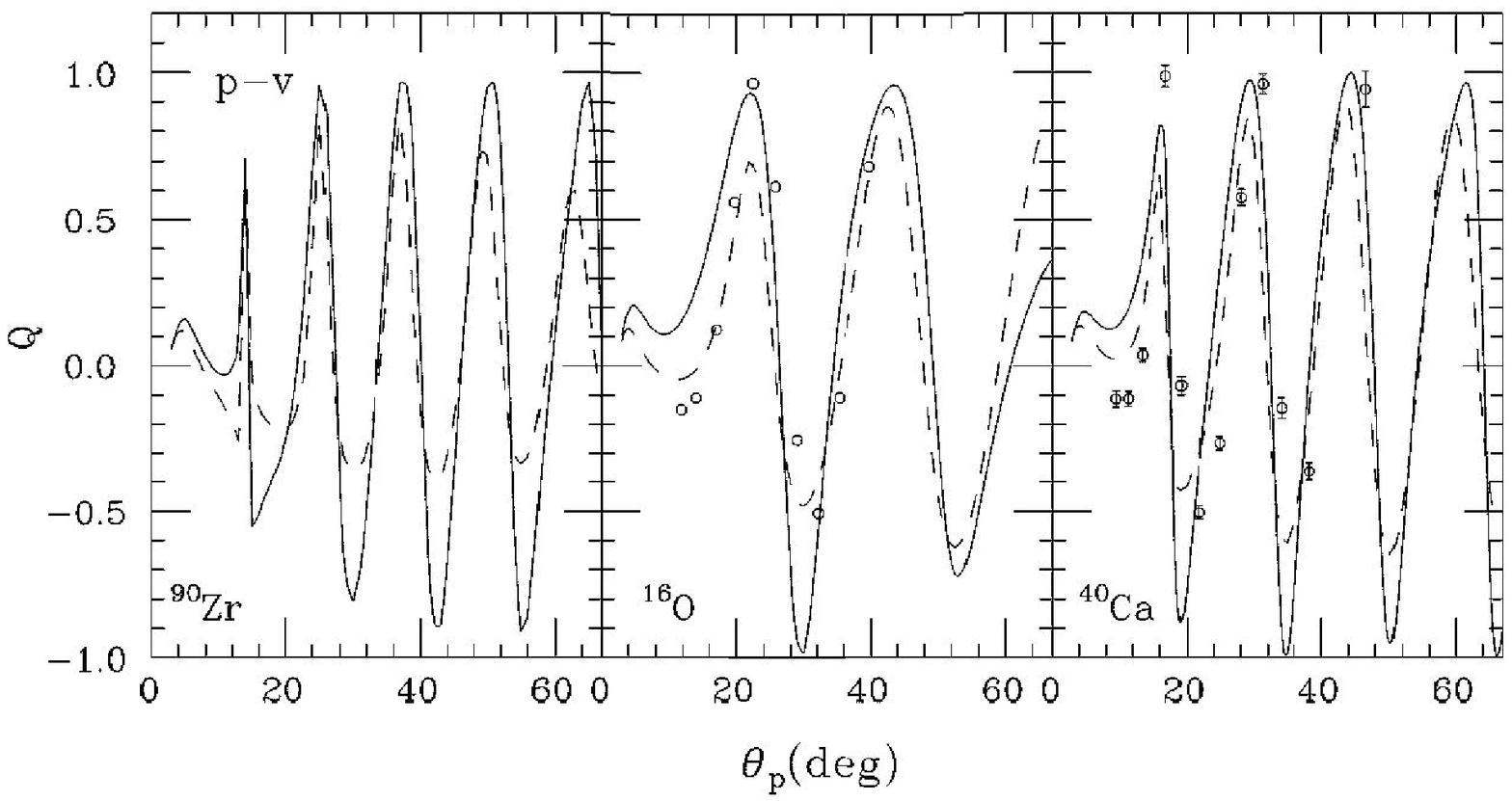}
\caption{\label{Fig7} Spin rotation matrix. Solid (dashed) lines are from calculations with approximate (exact) exchange and pseudovector  coupling. Data are from Ref. \cite{S85} and \cite{S82} as given in \cite{M87}.}
\end{figure*}
The cross sections, analyzing  power, and spin rotation matrix for elastic scattering of  protons from $^{90}$Zr, $^{16}$O, and $^{40}$Ca are shown in Figs. \ref{Fig2}-\ref{Fig7}, calculated with the RFL $t$ matrices of Ref. \cite{H85}.  Calculations were made for both the pseudoscalar and pseudeovector $\pi - N$ coupling.  The RFL parameters are those recommended in Ref. \cite{HMS91}.   No medium corrections were applied to the $t$ matrices.  The solid lines in these figures correspond to calculations with the exchange contribution approximated by the method of Ref. \cite{BR77}.  This converts the non-local potential to a local one which is generated by the program FOLDER \cite{HMS91}.  The dashed lines in these figures correspond to calculations with the exchange contribution calculated with Eq. (\ref{Eq22}), and hence, one has solved the non-local potential problem.  Looking at Figs. \ref{Fig2}, \ref{Fig4}, and \ref{Fig6}, one can conclude that calculating the exchange term exactly improves the agreements with data \cite{S85,S82}, except, perhaps for the $^{16}$O analyzing power.   For pseudovector coupling in Figs. \ref{Fig3}, \ref{Fig5}, and \ref{Fig7}, it is difficult to tell whether the approximate or exact exchange give better agreement with data.  However, the point is that the approximate and exact exchange give different results.  This is somewhat surprising in that the Breiva and Rook approximation has done reasonably well in nonrelativistic calculations.

\begin{figure}[!htbp]
\includegraphics[width=8cm]{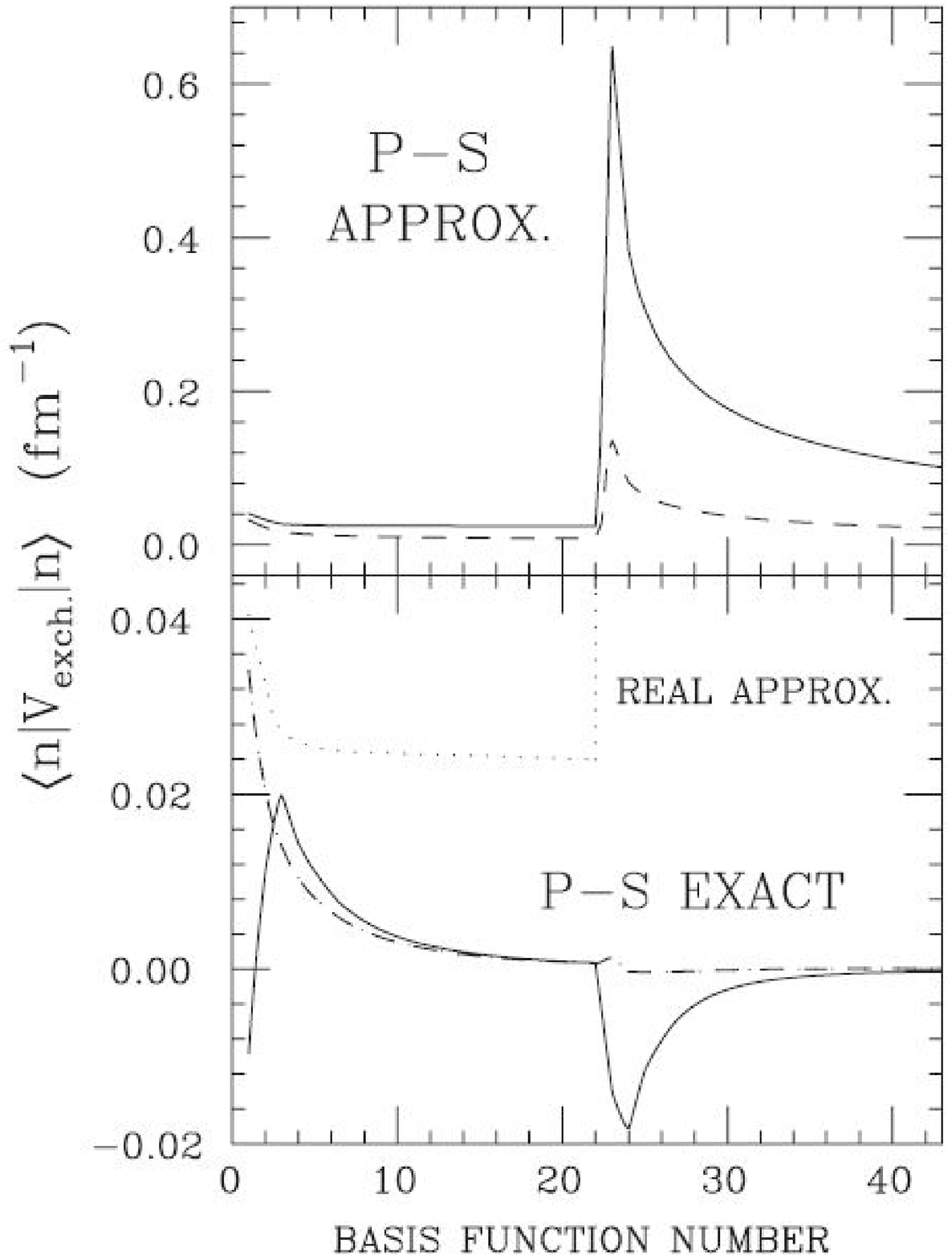}
\caption{\label{Fig8} Diagonal matrix elements of the exchange potential with pseudoscalar coupling versus basis function number for $\kappa = - 1$ Dirac oscillators.  Basis numbers $1$ to $22$ correspond to $n = 0$ to $21$ positive energy states, and basis numbers $23$ to $43$ correspond to $n = 1$ to $21$ negative energy states.  The upper panel is for the approximate exchange; the lower for the exact exchange. The solid line corresponds to the real part, the dashed to the imaginary. The dotted line repeats the approximate real values in the lower panel.}
\end{figure}
\begin{figure}[!htbp]
\includegraphics[width=8cm]{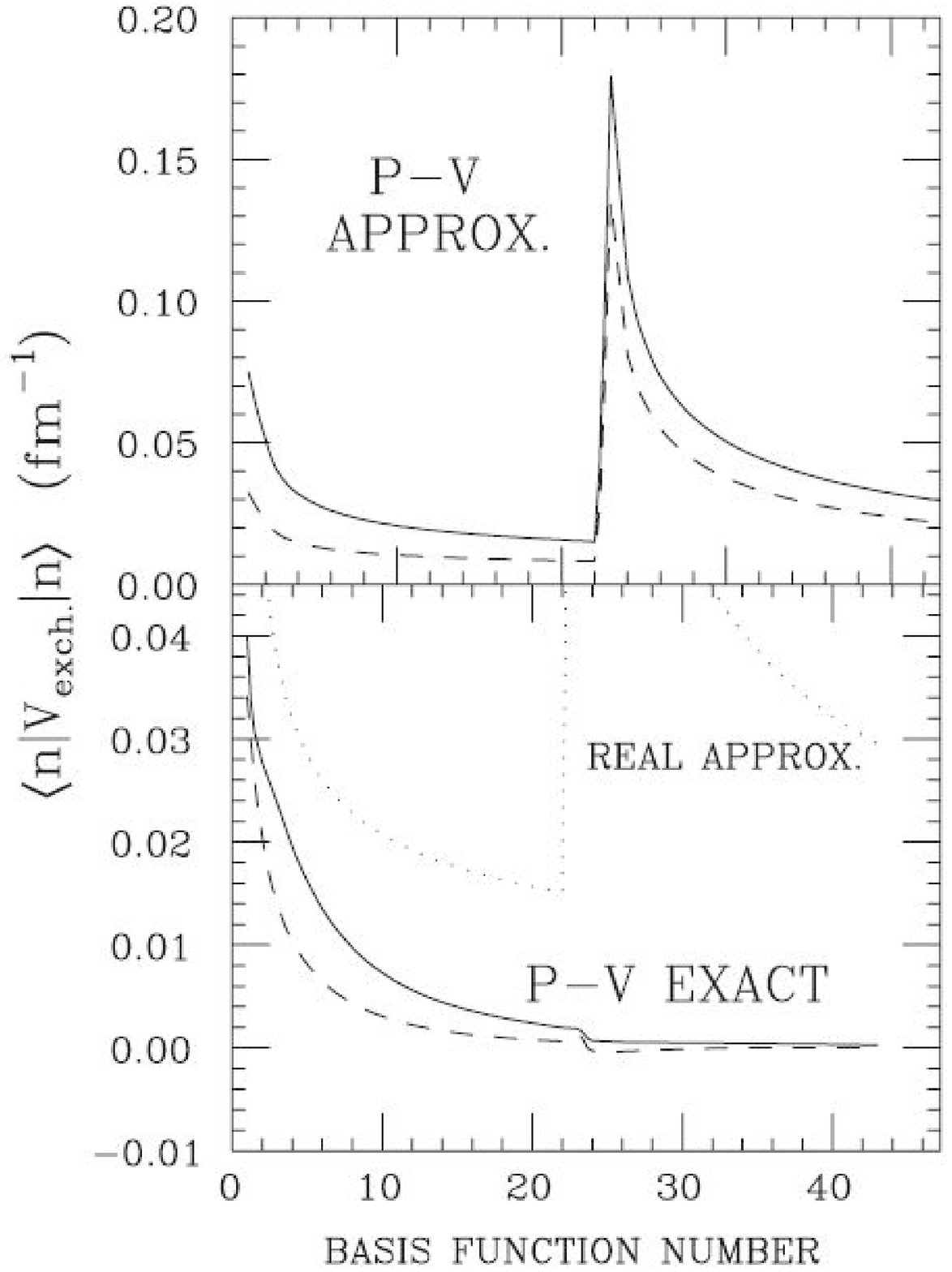}
\caption{\label{Fig9} Same as Fig. \ref{Fig8} but for pseudovector coupling.}
\end{figure}
Unfortunately, it is not possible to compare approximate and exact exchange optical potentials.  One can, however, compare the matrix elements of the Dirac oscillators for both the approximate and exact exchange.  The diagonal matrix elements of the exchange term for $\kappa = -1$ Dirac oscillators are shown in Figs. \ref{Fig8} and \ref{Fig9}.  The basis numbers 1 to 22 correspond to $n = 0$ to 21 positive energy states, and basis numbers 23 to 43 correspond to $n = 1$ to 21 negative energy states.  The upper panel is for the approximate exchange, the lower for the exact.  The solid line corresponds to the real part, the dashed to the imaginary.  The dotted line is also the approximate real values in the lower panel for comparison.  For the positive energy states the approximate matrix elements tend to be larger, perhaps an average of 1.7 larger.  However, for the negative energy states, the approximate matrix elements are very much larger and have the wrong sign.

\begin{figure}[!htbp]
\includegraphics[width=8cm]{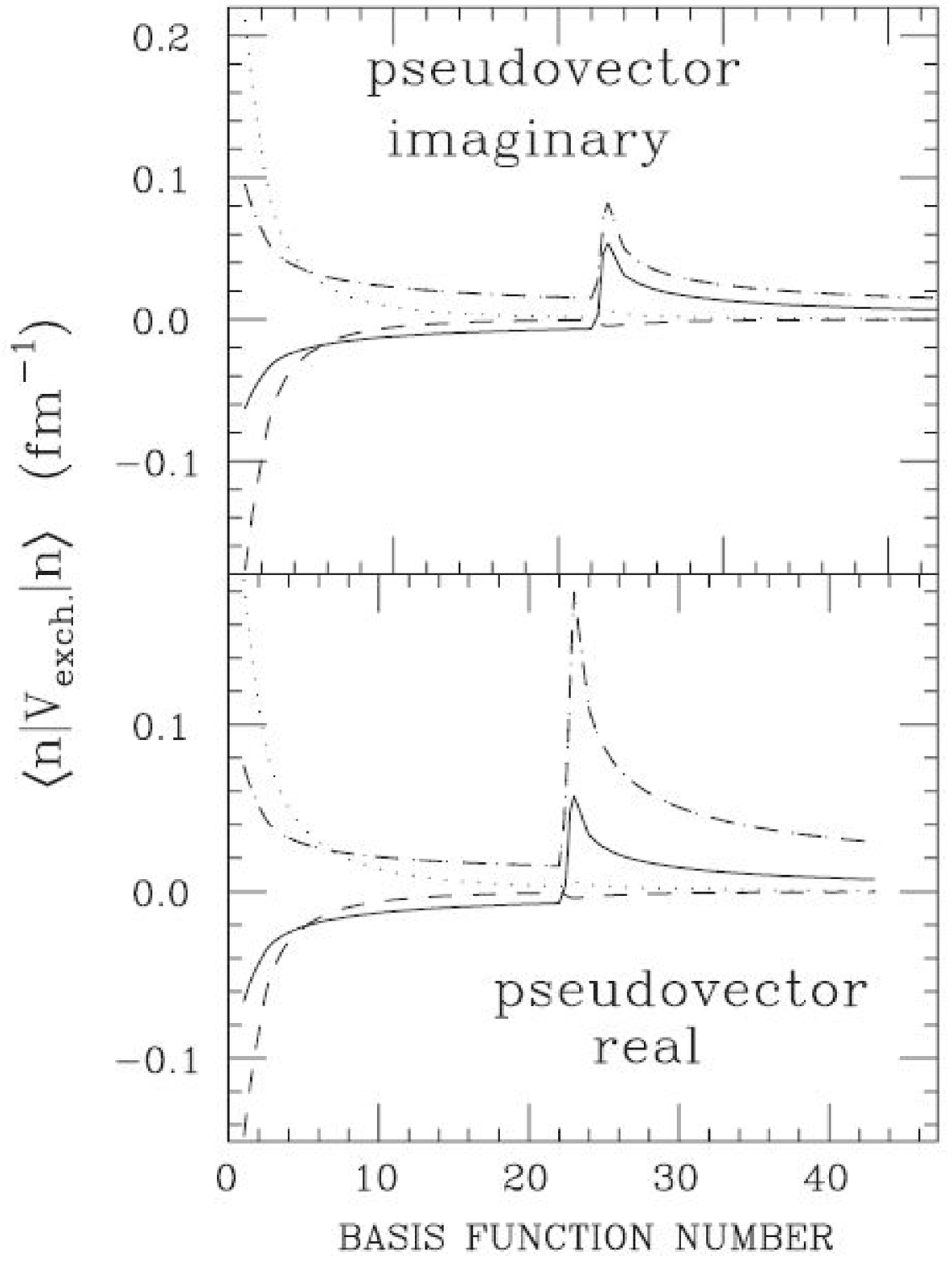}
\caption{\label{Fig10} Diagonal matrix elements of the exchange potential with pseudovector coupling versus basis function number for $\kappa = - 1$ Dirac oscillators.  The dashed (dotted) line corresponds to the contribution to the exact matrix elements originating from the scalar (vector) component of the $t$ matrix; the solid (dot--dashed) line corresponds to the contribution to the approximate matrix elements originating from the scalar (vector) component of the $t$ matrix.}
\end{figure}
The contributions to the pseudovector matrix elements are broken down further in Fig. \ref{Fig10}.  The dashed line corresponds to the contribution to the exact matrix elements originating from the scalar component of the $t$ matrix; the dotted line corresponds to the contribution to the exact matrix elements originating from the vector component of the $t$ matrix; the solid line corresponds to the contribution to the approximate matrix elements originating from the scalar component of the $t$ matrix; the dot-dashed line corresponds to the contribution to the approximate matrix elements originating from the vector component of the $t$ matrix.  One sees that the matrix elements between positive energy states are, on average, very similar when calculated with exact and approximate exchange.  This would be consistent with the Brieva and Rook approximation's working well in nonrelativistic calculations.  However, the approximate matrix elements between negative energy states are far too large, and both the scalar and vector contributions have the same sign.  The result is a significant difference in the observables, even for the RFL amplitudes which do not rely on sensitive cancellation between direct and exchange terms.

\begin{figure}[!htbp]
\includegraphics[width=8cm]{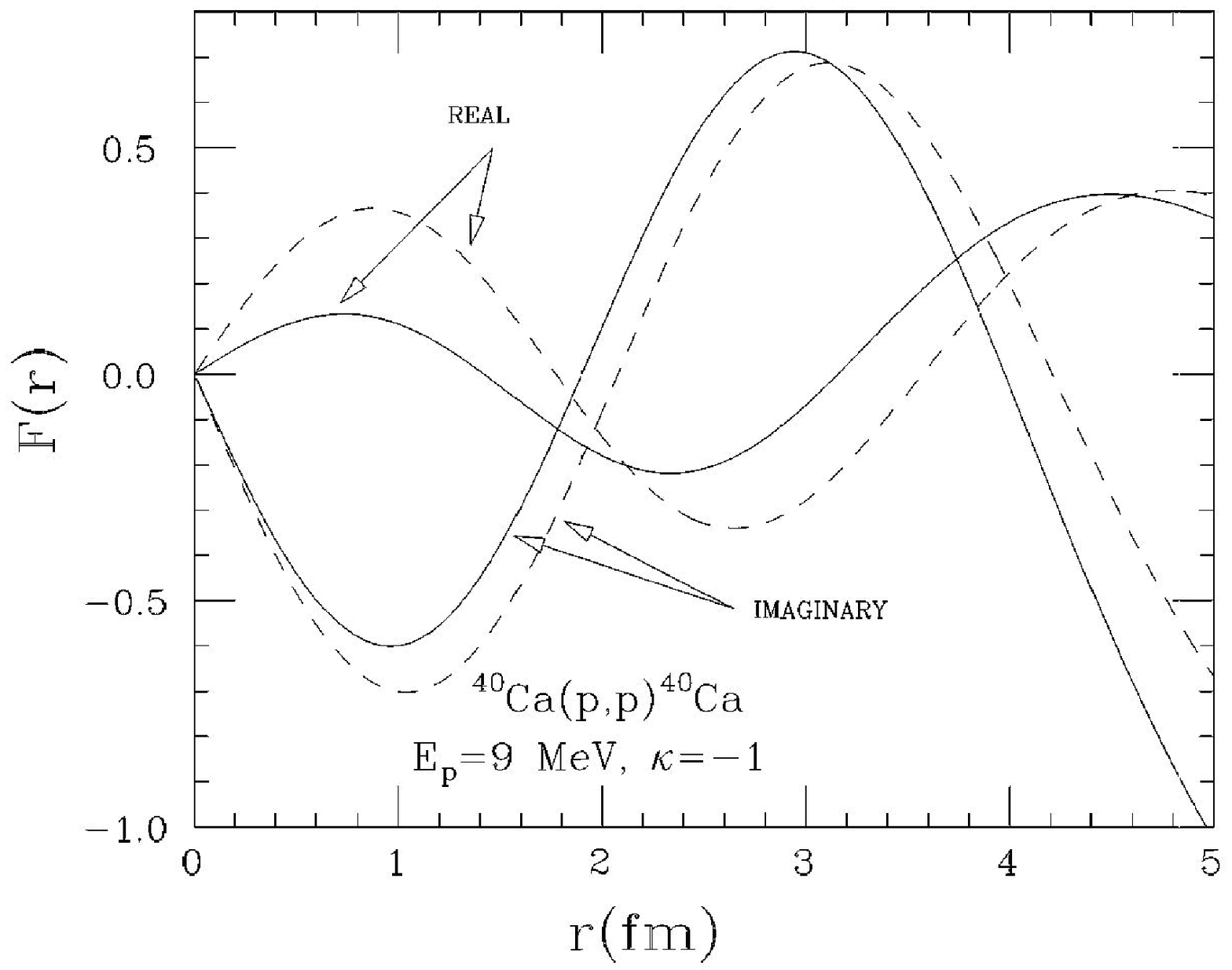}
\caption{\label{Fig11} $E_{p} = 9.0$ MeV, $\kappa = -1$, $p + ^{40}$Ca, generated with the potential from Ref. \cite{A81}. The dashed (solid) lines are upper components of the wave function with (without) projecting the $RMFT$ wave functions out of expansion basis.}
\end{figure}
Another advantage of the $R$-matrix approach is the ability to make the scattering states orthogonal to the bound states.  This would be particularly important if one were doing low energy capture reactions.  This is demonstrated in Fig. \ref{Fig11} with the $E_{p}= 9.0$ MeV, $\kappa = -1$, $p + ^{40}$Ca wave functions, generated with the potential from Ref. \cite{A81}.  The dashed (solid) lines are the upper components of the wave function with (without) projecting the RMFT wave functions out of the expansion basis.  

\begin{figure}[!htbp]
\includegraphics[width=8cm]{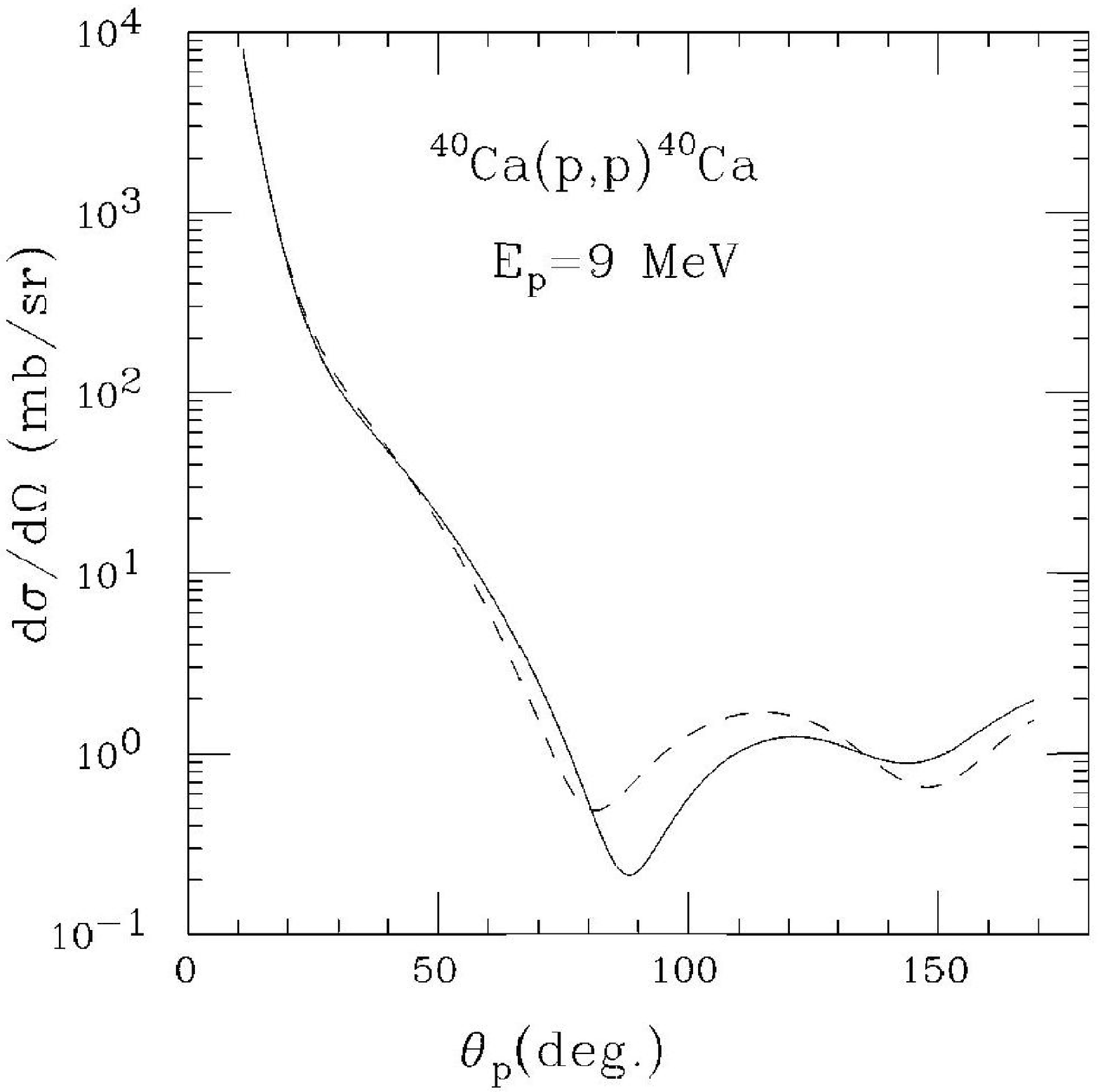}
\caption{\label{Fig12} Cross section plot of the $E_{p} = 9.0$ MeV, $p + ^{40}$Ca. The dashed (solid) lines is the cross section with (without) projecting the $RMFT$ wave functions out of expansion basis.}
\end{figure}

The cross section is also affected by this projection as shown in Fig. \ref{Fig12}.  The dashed line corresponds to projecting out the bound states which makes the nucleus look larger and more diffuse.  The effect of projecting out the bound states decreases as the energy increases, and by 50 MeV, it is not significant.
\FloatBarrier 

\section{Conclusions}

Simple expressions for the Dirac oscillators have been presented, and a review of the one-channel $R$-matrix approach to the Dirac equation given.  The formalism described above is easily extended to the multi-channel case, and as in the non-relativistic problem, many channels can be included.  The utility of employing Dirac oscillators as an expansion basis in the $R$-matrix formalism was demonstrated by performing RIA calculations for calculated for protons on $^{16}$O, and $^{40}$Ca and $^{90}$Zr.  In the $R$-matrix approach the exchange terms may be calculated exactly.  It was found that even for the $t$ matrices of Ref.\cite{H85}, which were deliberately constructed so as to not have sensitive cancellations between exchange and direct terms, a significant difference was found between observables calculated with the exact exchange and the plane-wave approximation for the exchange density.  Calculations employing the exact exchange improved agreement with data for pseudoscalar $\pi - N$ coupling in the RFL amplitudes, but for pseudovector coupling the agreement was similar to that of calculations with the approximate exchange.  The differences between the exact and approximate exchange were traced back to the matrix elements between negative energy states of the expansion basis, and are, therefore, relativistic in origin.

\section*{Acknowledgments}
This work was supported by the National Science Foundation under grant PHY-0456943.

\end{document}